\documentclass[aps,prb,twocolumn,showpacs,superscriptaddress]{revtex4}
\usepackage{graphicx}

\begin{document}
\title {Long-Range Magnetic Interactions in the Multiferroic Antiferromagnet
$\rm MnWO_4$ }

\author{Feng Ye}
\email{yef1@ornl.gov}
\affiliation{Neutron Scattering Science Division,
Oak Ridge National Laboratory, Oak Ridge, Tennessee 37831, USA }

\author{Randy S.~Fishman}
\affiliation{Materials Science and Technology Division, Oak Ridge National
Laboratory, Oak Ridge, Tennessee 37831, USA}

\author{Jaime A.~Fernandez-Baca}
\affiliation{Neutron Scattering Science Division,
Oak Ridge National Laboratory, Oak Ridge, Tennessee 37831, USA }
\affiliation{Department of Physics and Astronomy,
The University of Tennessee, Knoxville, Tennessee 37996, USA}

\author{Andrey A. Podlesnyak}
\affiliation{Neutron Scattering Science Division,
Oak Ridge National Laboratory, Oak Ridge, Tennessee 37831, USA }

\author{Georg Ehlers}
\affiliation{Neutron Scattering Science Division,
Oak Ridge National Laboratory, Oak Ridge, Tennessee 37831, USA }

\author{Herbert A.~Mook}
\affiliation{Neutron Scattering Science Division,
Oak Ridge National Laboratory, Oak Ridge, Tennessee 37831, USA }

\author{Yaqi Wang}
\affiliation{Department of Physics and TCSUH, University of
Houston, Houston, Texas 77204, USA}

\author{Bernd~Lorenz}
\affiliation{Department of Physics and TCSUH, University of
Houston, Houston, Texas 77204, USA}

\author{C.~W.~Chu}
\affiliation{Department of Physics and TCSUH, University of
Houston, Houston, Texas 77204, USA}

\date{\today}

\begin{abstract}
The spin-wave excitations of the multiferroic $\rm MnWO_4$ have been measured
in its low-temperature collinear commensurate phase using high-resolution
inelastic neutron scattering. These excitations can be well described by a
Heisenberg model with competing long-range exchange interactions and a
single-ion anisotropy term. The magnetic interactions are strongly frustrated
within the zigzag spin chain along $c$ axis and between chains along the
$a$ axis, while the coupling between spin along the $b$ axis is much weaker.
The balance of these interactions results in the noncollinear incommensurate
spin structure associated with the magnetoelectric effect, and the perturbation
of the magnetic interactions leads to the observed rich phase diagrams of the
chemically-doped materials. This delicate balance can also be tuned by the
application of external electric or magnetic fields to achieve magnetoelectric
control of this type of materials.
\end{abstract}

\maketitle

Magnetoelectric multiferroic materials, which exhibit the coexistence of
ferroelectric (FE) and magnetic orders, have attracted great attention in
recent years.\cite{tokura06,khomskii06,cheong07,kimura07} Several classes of
multiferroics among transition metal oxides have been discovered including
geometrically-frustrated Cu$M$O$_2$ ($M$ is Fe, Cr), \cite{kimura06,kimura08}
RbFe(MoO$_4$)$_2$ \cite{kenzelmann07}, $\rm Ni_3V_2O_8$ \cite{lawes05}, or
rare-earth ($R$) manganites $R$MnO$_3$ and $R$$\rm
Mn_2O_5$.\cite{kimura03,goto04,hur04} The ability to simultaneously control the
electric (E) and magnetic (M) properties makes these multiferroics promising
candidates for technological applications.\cite{kimura03,yamasaki07}
A characteristic feature in those magnetically induced multiferroics
is the presence of long-range magnetic structures with noncollinear spiral spin
configurations.  Such magnetic order is a consequence of magnetic frustration
either due to geometric constraints or competing exchange interactions
resulting in a close competition of different magnetic structures that are
nearly degenerate in energy.

The mineral H\"{u}bnerite MnWO$_4$ appears to be an unique material that
not only exhibits intriguing multiferroic phenomena but also shows
rich magnetic phases via chemical substitutions.
\cite{taniguchi06,heyer06,arkenbout06,sagayama08,ye08,meddar09,song09,chaudhury11,finger10}
It has been considered one of the prototypical multiferroics capable
of magnetoelectric (ME) control.\cite{taniguchi08,sagayama08} Unlike
$R$MnO$_3$ where the spiral magnetic structure often involves
ordering of the rare-earth moments, $\rm MnWO_4$ is a frustrated
antiferromagnet (AF) with only one type of magnetic ion.  The
Mn$^{2+}$ spins ($S$=5/2) undergo successive transitions in zero
field.\cite{lautenschlager93} The low-temperature ($T$) magnetic
structure has a collinear spin configuration [Fig.~1(a)].  For $T$
between 7.8~K ($T_{N1}$) and 12~K ($T_{N2}$), the magnetic structure
evolves into an incommensurate (ICM) elliptical spiral configuration
accompanied by a spontaneous electric polarization $P$ along the
crystalline $b$ axis. When $T$ is further raised between $T_{N2}$
and $T_{N3} (\approx 13.5$~K), MnWO$_4$ becomes a collinear ICM and
paraelectric.

In MnWO$_4$, the electric polarization that is correlated with the
spiral magnetic structure can be well understood by the microscopic
picture regarded as inverse Dzyaloshinskii-Moriya interaction.
\cite{katsura05,sergienko06,mostovoy06} On the other hand,
characterizing the magnetic interactions that cause the formation of
the complex spin structures and understanding how the modification
of exchange couplings affects the evolution between different
phases remains an unresolved issue despite intensive experimental
and theoretical studies.\cite{ehrenberg99,tian09,hollmann10} Early
inelastic neutron scattering (INS) work suggested that the stabilization
of the collinear configuration requires higher order magnetic
interactions.\cite{ehrenberg99} Although the magnetic exchange
parameters obtained by Ehrenberg {\it et al.}~fit the experimental
data, the longest bond distance was associated with the strongest
coupling constant.  Later, density functional calculation and
classical spin analysis were performed to investigate the magnetic
structure and FE polarization in MnWO$_4$.\cite{tian09}  Tian {\it
et al.}~concluded that the spin exchange interactions are frustrated
along both the $a$ and $c$ axes. However, a quantitative
experimental characterization of the magnetic interactions is still
lacking.  Here we report high-resolution INS measurements that
show that the low-$T$ magnetic ground state of pure MnWO$_4$ indeed
results from the competition of long-range interactions that are
highly frustrated and sensitive to small perturbations. The
comprehensive mapping of the of the magnetic excitations along
several symmetry directions allowed an unambiguous determination of
the dispersion relations and the exchange interactions. Most
importantly, such microscopic characterization of the spin coupling
constants provides a fundamental step toward the construction of the
ground state Hamiltonian from which the FE phase can be derived.

\begin{figure}[ht!]
\includegraphics[width=2.6in]{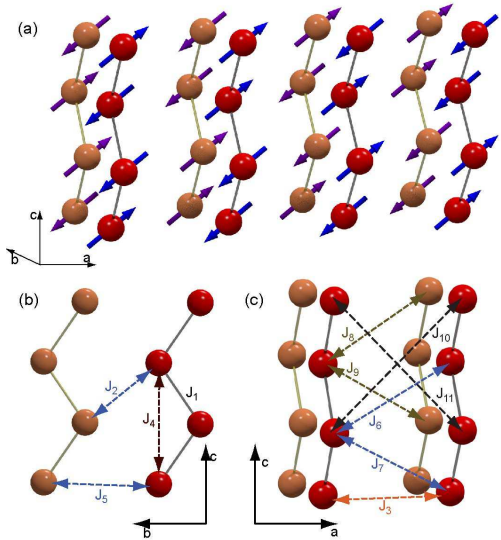}
\caption{\label{fig:structure} (Color online) (a) The
magnetic structure of $\rm MnWO_4$ in the collinear, commensurate
phase at low temperature. The magnetic spins lie in the $ac$ plane
with the moment canted to the $a$ axis about 35$^o$.
\cite{lautenschlager93} The magnetic spins form zigzag $\uparrow
\uparrow \downarrow \downarrow$ chains along the $c$ axis, and are
coupled antiferromagnetically along the $b$ axis. (b) The magnetic
interactions along and between spin chains in the $bc$ plane. (c)
Higher order magnetic interactions along the $a$ axis direction.  The
magnetic couplings are labeled with increasing bonding distance.
Note the monoclinic crystal structure ($\beta=91.14^\circ$) makes
$J_6$/$J_7$, $J_8$/$J_9$, $J_{10}$/$J_{11}$ pairs different.
}
\end{figure}

A 5-g single crystal of $\rm MnWO_4$ was grown by the floating-zone
technique. Neutron diffraction was performed on a small piece of this crystal
(0.2 g) to verify the spin structure using the four-circle single crystal
diffractometer HB3A at the High Flux Isotope Reactor at the Oak Ridge National
Laboratory (ORNL).  The inelastic neutron scattering measurements were
performed using the Cold Neutron Chopper Spectrometer (CNCS) at the Spallation
Neutron Source at ORNL.  The momentum transfer wavevectors $q=(q_x,q_y,q_z)$
are in units of $\rm \AA^{-1}$ at positions
$(H,K,L)=(q_xa/2\pi,q_yb/2\pi,q_zc/2\pi)$ in reciprocal lattice units (rlu),
where $a=4.83~\AA$, $b=5.75~\AA$, $c=4.99~\AA$. We aligned the crystal in
several scattering planes such that the spin-wave (SW) dispersion along the
[1,0,-2], [1,0,2] and [0,1,0] symmetric directions that pass across the
magnetic Bragg peaks can be readily measured. The incident neutrons with
wavelength of $\lambda=4.4~\AA$ were chosen to ensure the needed energy
resolution to separate various magnetic excitation branches.

\begin{figure}[ht!]
\includegraphics[width=3.2in]{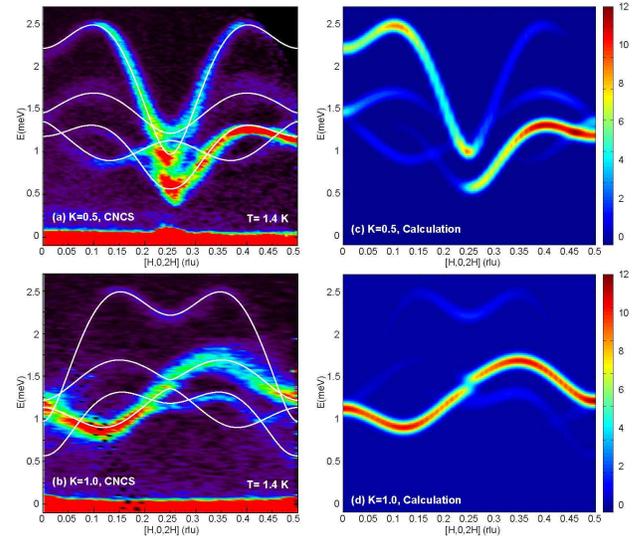}
\caption{\label{fig:scanh0lp} (Color online)
(a) SW dispersion spectra along the [1,0,2] direction
through the magnetic peak (1/4,1/2,1/2). (b) Magnetic excitation
spectra along the same direction with $K=1.0$. (c and d) Calculated
spectra along [1,0,2] direction with $K=0.5$ and $K=1.0$ using
magnetic exchange interactions up to $J_{11}$ with instrument
resolution convoluted. The solid lines overlapped with
experimental data in (a) and (b) are the predicted dispersion curves.
}
\end{figure}

MnWO$_4$ orders in the collinear phase with two inequivalent
commensurate wavevectors $q_M=(1/4,1/2,\pm1/2)$. Figures 2(a) and 2(b)
present the spin excitation spectra along the [1,0,2] direction with
$K=0.5$ and $1.0$. This scanning direction goes through the magnetic
Bragg peak $(1/4,1/2,1/2)$. The data clearly show four distinct
branches that disperse out from the magnetic zone center (ZC) to the
zone boundary. The spectra reveal a spin gap of 0.5~meV and
boundary energy around 2.2~meV. The excitation bandwidth is
consistent with the energy scale of the ordering temperature of
13.5~K.  The intensity of the excitation spectra is highly
asymmetric with respect to the magnetic Bragg point. For example,
the spectral weight of lowest branch in Fig.~2(a) is completely
missing as $H$ approaches zero, while it shows the highest intensity
as it moves toward $H=0.5$. This highlights the importance of
a complete survey in reciprocal space to fully map out the magnetic
dynamics.

Figures 3(a) and 3(b) display the magnetic scattering spectra along the
[1,0,-2] direction that crosses the other magnetic wavevector
$(1/4,1/2,-1/2)$, while Figs.~4(a) and 4(b) illustrate the spectra along
the [0,1,0] direction. The icattering intensity map shows similar
asymmetric feature on both sides of the magnetic ZC.
Investigations along those two directions have been previously
reported,\cite{ehrenberg99} but the lack of spectral weight for certain
branches makes a correct description of the SW dispersion difficult.

\begin{table*}[ht!]
    \caption{Magnetic exchange coupling parameters derived from spin
    wave model calculation according to the Eq.~1. The Mn$^{2+}$
    ions located at position $(1/2,0.685,1/4)$ interact with
    neighboring spins through one or two oxygens. The bonding
    distance (in units of $\AA$) between Mn$\ldots$Mn are also
    listed. The magnetic interaction constants from previous
    experimental and theoretical studies have been normalized in
    units of meV for comparison.
    }
    \label{tab1}
\begin{tabular}{lrrrrrrrrrrrrrr}
\hline \hline
 & $J_1$  & $J_2$ & $J_3$ & $J_4$  & $J_5$ & $J_6$ & $J_7$ & $J_8$ & $J_9$ & $J_{10}$  & $J_{11}$ & $D$ & $\chi^2$\\
    \hline
Mn$\ldots$Mn & $3.286$ & $4.398$ & $4.830$ & $4.990$ & $5.760$ & $5.801$ & $5.883$ & $6.496$ & $6.569$ & $6.875$ & $7.013$ &  & \\
This work & $-0.47(1)$ & $-0.05(1)$ & $-0.48(1)$ & $-0.21(1)$ & $0.09(1)$ & $-0.49(1)$ & $-0.12(1)$ & $0.05(1)$ & $-0.23(1)$ & $--$ & $--$ & $0.12(1)$ & $2.62$ \\
This work & $-0.42(1)$ & $-0.04(1)$ & $-0.32(1)$ & $-0.26(1)$ & $0.05(1)$ & $-0.43(1)$ & $-0.12(1)$ & $0.02(1)$ & $-0.26(1)$ & $-0.15(1)$ & $0.02(1)$ & $0.09(1)$ & $1.11$\\
Ref.~\onlinecite{ehrenberg99} & $-0.084$ & $-0.058$ & $-0.182$ & $0.178$ & $0.009$ & $-0.219$ & $0.010$ & $0.212$ & $-0.980$ & $--$ & $--$ & $0.061$ & $--$\\
Ref.~\onlinecite{tian09} & $-0.160$ & $-0.016$ & $-0.153$ & $-0.232$ & $-0.018$ & $-0.089$ & $-0.185$ & $-0.031$ & $-0.115$ & $--$ & $--$ & $--$ & $--$\\
    \hline \hline
    \end{tabular}
\end{table*}

\begin{figure}[th!]
\includegraphics[width=3.2in]{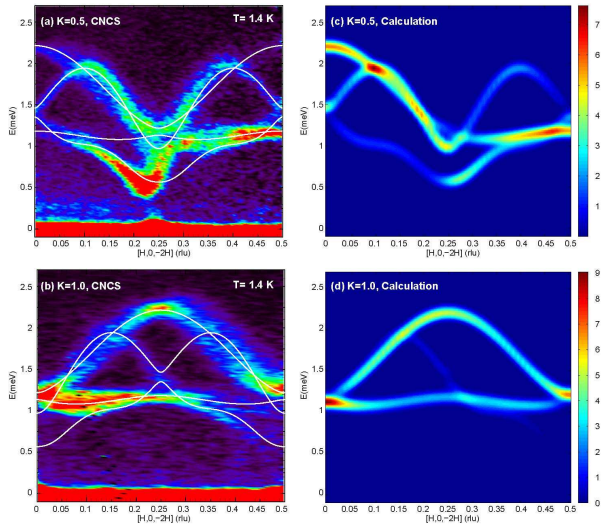}
\caption{\label{fig:scanh0ln} (Color online)
(a) SW dispersion spectra along the [1,0,-2] direction through the
magnetic peak (1/4,1/2,-1/2). (b) Magnetic excitation spectra
along the same direction with $K=1.0$. (c)-(d) Calculated spectra along
the same symmetric directions with $K=0.5$ and $K=1.0$.
}
\end{figure}

\begin{figure}[ht!]
\includegraphics[width=3.2in]{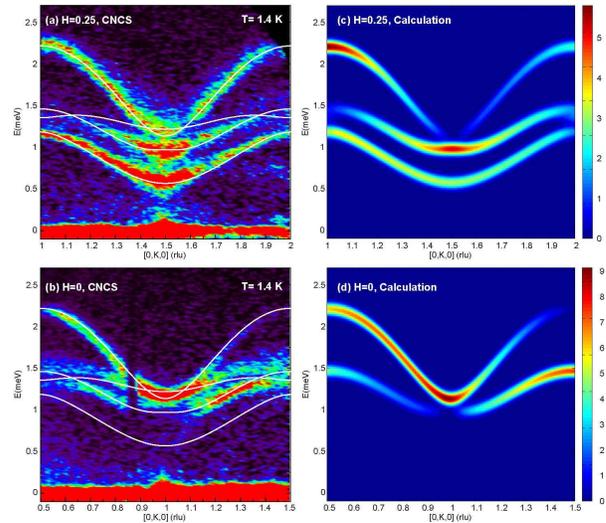}
\caption{\label{fig:scan0k0} (Color online)
(a) SW dispersion spectra along the [0,1,0] direction through the
magnetic peak (1/4,1/2,-1/2). (b) Magnetic excitation spectra
along the same direction through the nuclear peak (0,1,0).  (c)-(d)
Calculated spectra along the same symmetric directions with $H=0.25$
and $H=0$.
}
\end{figure}

The SW dispersion curves can be modeled by a general effective
Heisenberg Hamiltonian:
\begin{equation}
H=-\sum_{i,j} J_{i,j} {\bf{S}}_i\cdot{\bf{S}}_j - D\sum_{i} S^{2}_{iz},
\end{equation}
where $\sum_{i,j}$ indicates summation over pairs of spins,
\cite{spinexch} $D$ is the single-ion anisotropy, and $S_{iz}$
denote the spin components along the easy axis.  To
calculate the corresponding spectral weight, the low-$T$ spin
structure has been verified by collecting a complete set of magnetic
reflections that covers the entire reciprocal space. The spin
configuration obtained by Rietveld refinement with group theory
analysis is in good agreement with a previous report.
\cite{lautenschlager93}  This configuration is then used to
evaluate the magnetic scattering cross-section:

\begin{equation}
\frac{d^2\sigma}{d\Omega dE} \propto  f^2(Q)
e^{-2W}\sum_{\alpha\beta}(\delta_{\alpha\beta}-\hat{Q}_{\alpha}\hat{Q}_{\beta})
S^{\alpha\beta}({\bf{Q}},\omega)
\end{equation}
where $f^2(Q)$ is the magnetic form factor for
the Mn$^{2+}$ ion, $e^{-2W}$ is the Debye-Waller factor, $\hat{Q}_\alpha$ is
the $\alpha$ component of a unit vector in the direction of $\bf Q$,
and $S^{\alpha\beta}({\bf{Q}},\omega)$ is the response function that
describes the $\alpha\beta$ spin-spin correlations.

Combining data along all symmetry directions, the dispersion
relations can be simultaneously modeled using Eq.~(1). We started
with nine exchange parameters and single-ion anisotropy $D$, as
assumed in Ref.~\onlinecite{ehrenberg99}. Although the fitting
parameters provide a fair description of the data along the [1,0,-2]
and [0,1,0] direction, they fail to capture the dispersion relations
along the [1,0,2] direction. Instead, an extra pair of magnetic
interactions $J_{10}$ and $J_{11}$ have to be included to achieve
satisfactory agreement for the data along all scan directions. As
listed in Table~I, the magnetic exchange constants generally
decrease in amplitude as the bonding distances increase, except the
weaker $J_2$ and $J_5$ (both close to zero) along the $b$ axis.
Mn$^{2+}$ has an electronic configuration of $3d^5$ (orbital singlet)
that is not expected to have any magnetic anisotropy, and the
nonvanishing $D=0.09$~meV indicates a possible spin-orbit coupling
that causes the pinning of the magnetic moments in the $ac$ plane.
\cite{hollmann10} To test whether those parameters are consistent
with the actual spin structure, the magnetic energy for all
possible spin configurations ($2^8=256$ with eight spins in one
magnetic unit cell) are calculated.  We verified that only the spin
structure depicted in Fig.~1(a) gives the lowest energy. In
addition, we employed the spin-rotation technique to calculate
the expected intensity for the SW modes \cite{haraldsen09} and to
compare with the observed wavevector dependence of the spectra
weight. The right panels of Figs.~2-4 show the excitation spectra
map using Eq.~(2) with fitted parameters.  The excellent agreement
between the calculated intensities and the experimental data
provides convincing evidence that the formation of the collinear
spin structure indeed requires longer range magnetic interactions.

A complex spin configuration would form in the conventional 1D
frustrated spin chain if the next-nearest-neighbor antiferromagnetic (AFM)
interaction $J'$ becomes substantially stronger compared to the
nearest-neighbor (NN) ferromagnetic (FM) exchange coupling $J$. A
spiral phase might appear for $J'/|J|>1/4$ and a collinear $\uparrow
\uparrow \downarrow \downarrow$ structure will emerge when
$J'/|J|>1/2$.\cite{cheong07} The zigzag $E$-type phase observed in
$R$MnO$_3$ is a classic example of competing short-range FM-AFM
interactions caused by lattice distortions.\cite{kimura03b,mochizuki10}
In the case the MnWO$_4$, the exchange
coupling are predominantly AFM and three-dimensional.  Nevertheless,
strong long-range magnetic interactions comparable to NN interaction
($J_1$) along the $c$ axis ($|J_4/J_1|>1/2$) and $a$ axis
($|J_3/J_1|>1/2$, $|J_6/J_1| \approx 1$ and $|J_7/J_1|>1/4) $ are
observed. This reflects the magnetic frustration in those two
directions and is consistent with the ICM components present in the $a$
and $c$ directions but not in the $b$ direction, when the system enters the
spiral phase.  The exchange coupling remains sizable
($J_{10}$) even at a rather long distance. Such an unusual extended
interaction could also be viewed as a much-reduced NN exchange
coupling because of a nearly 90$^\circ$ Mn-O-Mn bonding angle.
\cite{goodenough63} Thus, the unique crystalline structure makes
MnWO$_4$ a promising material to achieve novel physical properties
when the magnetic interactions are modified. It was reported that
doping a few percent magnetic or nonmagnetic impurities can
drastically affect the spin order.\cite{ye08,meddar09,song09,chaudhury11}
For instance, while replacing Mn$^{2+}$ with Fe$^{2+}$ ions that
have a larger local magnetic anisotropy seems to enhance the
collinear structure,\cite{ye08} the introduction of nonmagnetic
Zn$^{2+}$ ions that weaken the overall magnetic interactions
switches the ground state from collinear to spiral
order.\cite{meddar09,chaudhury11} Those results demonstrate that
chemical substitutions are a viable tool to tune the multiferroic
properties in the extremely sensitive MnWO$_4$.

In summary, high-resolution INS is used to study the SWs in the
collinear phase of MnWO$_4$. The collinear spin order is stabilized
by the delicate balance of competing long-range magnetic
interactions. We provide an effective Hamiltonian to describe the
highly frustrated magnetic order within the zigzag spin chains along
the $c$ axis and between spin chains along $a$ axis.  Rich and complex
magnetic phases are expected in chemically doped MnWO$_4$ due to the
fine-tuning of the magnetic interactions.  The delicate balance of
exchange interactions in MnWO$_4$ can also be used to achieve
magnetoelectric control using external electric or magnetic fields.

This work was partially supported by the Division of Scientific User
Facilities of the Office of Basic Energy Sciences, U.S.~Department of
Energy. Work at Houston is supported in part by the T.L.L.  Temple
Foundation, the J. J. and R. Moores Endowment, and the state of
Texas through TCSUH and at LBNL through the US DOE, Contract No.
DE-AC03-76SF00098.


\begin{references}
\bibitem{tokura06}
Y.~Tokura,
\newblock Science {\bf 312}, 1481 (2006).

\bibitem{khomskii06}
D.~I. Khomskii,
\newblock J. Magn. Magn. Mater. {\bf 306}, 1 (2006).

\bibitem{cheong07}
S.-W. Cheong and M.~Mostovoy,
\newblock Nature Materials {\bf 6}, 13 (2007).

\bibitem{kimura07}
T.~Kimura,
\newblock Annu. Rev. Mater. Res. {\bf 37}, 387 (2007).

\bibitem{kimura06}
T.~Kimura, J.~C. Lashley, and A.~P. Ramirez,
\newblock Phys. Rev. B {\bf 73}, 220401(R) (2006).

\bibitem{kimura08}
K.~Kimura, H.~Nakamura, K.~Ohgushi, and T.~Kimura,
\newblock Phys. Rev. B {\bf 78}, 140401(R) (2008).

\bibitem{kenzelmann07}
M.~Kenzelmann {\em et~al.},
\newblock Phys. Rev. Lett. {\bf 98}, 267205 (2007).

\bibitem{lawes05}
G.~Lawes {\em et~al.},
\newblock Phys. Rev. Lett. {\bf 95}, 087205 (2005).

\bibitem{kimura03}
Kimura {\em et~al.},
\newblock Nature (London) {\bf 426}, 55 (2005).

\bibitem{goto04}
T.~Goto, T.~Kimura, G.~Lawes, A.~P. Ramirez, and Y.~Tokura,
\newblock Phys. Rev. Lett. {\bf 92}, 257201 (2004).

\bibitem{hur04}
N.~Hur {\em et~al.},
\newblock Nature (Lodon) {\bf 429}, 392 (2004).

\bibitem{yamasaki07}
Y.~Yamasaki {\em et~al.},
\newblock Phys. Rev. Lett. {\bf 98}, 147204 (2007).

\bibitem{taniguchi06}
K.~Taniguchi, N.~Abe, T.~Takenobu, Y.~Iwasa, and T.~Arima,
\newblock Phys. Rev. Lett. {\bf 97}, 097203 (2006).

\bibitem{heyer06}
O.~Heyer {\em et~al.},
\newblock J. Phys.: Condens. Matter {\bf 18}, L471 (2006).

\bibitem{arkenbout06}
A.~H. Arkenbout, T.~T.~M. Palstra, T.~Siegrist, and T.~Kimura,
\newblock Phys. Rev. B {\bf 74}, 184431 (2006).

\bibitem{sagayama08}
H.~Sagayama {\em et~al.},
\newblock Phys. Rev. B {\bf 77}, 220407(R) (2008).

\bibitem{ye08}
F.~Ye {\em et~al.},
\newblock Phys. Rev. B {\bf 78}, 193101 (2008).

\bibitem{meddar09}
L.~Meddar {\em et~al.},
\newblock Chem. Mater. {\bf 21}, 5203 (2009).

\bibitem{song09}
Y.-S. Song, J.-H. Chung, J.~M.~S. Park, and Y.-N. Choi,
\newblock Phys. Rev. B {\bf 79}, 224415 (2009).

\bibitem{chaudhury11}
R.~P. Chaudhury {\em et~al.},
\newblock Phys. Rev. B {\bf 83}, 014401 (2011).

\bibitem{finger10}
T.~Finger {\em et~al.},
\newblock Phys. Rev. B {\bf 81}, 054430 (2010).

\bibitem{taniguchi08}
K.~Taniguchi, N.~Abe, H.~Umetsu, H.~A. Katori, and T.~Arima,
\newblock Phys. Rev. Lett. {\bf 101}, 207205 (2008).

\bibitem{lautenschlager93}
G.~Lautenschl{\"{a}}ger {\em et~al.},
\newblock Phys. Rev. B {\bf 48}, 6087 (1993).

\bibitem{katsura05}
H.~Katsura, N.~Nagaosa, and A.~V. Balatsky,
\newblock Phys. Rev. Lett. {\bf 95}, 057205 (2005).

\bibitem{sergienko06}
I.~A. Sergienko and E.~Dagotto,
\newblock Phys. Rev. B {\bf 73}, 094434 (2006).

\bibitem{mostovoy06}
M.~Mostovoy,
\newblock Phys. Rev. Lett. {\bf 96}, 067601 (2006).

\bibitem{ehrenberg99}
H.~Ehrenberg, H.~Weitzel, H.~Fuess, and B.~Hennion,
\newblock J. Phys.: Condens. Matter {\bf 11}, 2649 (1999).

\bibitem{tian09}
C.~Tian {\em et~al.},
\newblock Phys. Rev. B {\bf 80}, 104426 (2009).

\bibitem{hollmann10}
N.~Hollmann {\em et~al.},
\newblock Phys. Rev. B {\bf 82}, 184429 (2010).

\bibitem{spinexch} Spin pairs up to 11th NN are included to
calcuate the dispersion relation. This limits magnetic coupling
between Mn ions through zero or one WO$_6$ octahedron.

\bibitem{haraldsen09}
J.~T. Haraldsen and R.~S. Fishman,
\newblock J. Phys.: Condens. Matter {\bf 21}, 216001 (2009).

\bibitem{kimura03b}
T.~Kimura {\em et~al.},
\newblock Phys. Rev. B {\bf 68}, 060403(R) (2003).

\bibitem{mochizuki10}
M.~Mochizuki, N.~Furukawa, and N.~Nagaosa,
\newblock Phys. Rev. Lett. {\bf 105}, 037205 (2010).

\bibitem{goodenough63}
J.~B. Goodenough,
\newblock {\em Magnetism and Chemical Bond}, (Interscience, New York, 1963).
\end{references}
\end{document}